\documentclass[aps,twocolumn,superscriptaddress,prl,floatfix]{revtex4}
\usepackage{amssymb}
\usepackage{amsmath}
\usepackage{graphicx}

\def\Gq{e^2/h}
\def\Rq{h/2e^2}
\def\AlOx{\mathrm{Al}_2 \mathrm{O}_3}
\def\SiOx{\mathrm{SiO}_2}
\def\Hethree{^3\mathrm{He}}
\def\Kb{k_{B}}
\def\den{n_{s}}
\def\Te{T_{e}}
\def\Isd{I}
\def\Vsd{V_{\mathrm{sd}}}
\def\Vbg{V_{\mathrm{bg}}}
\def\Vtg{V_{\mathrm{tg}}}
\def\Si{S_{I}}
\def\Siex{S_{I}^{\mathrm{e}}}
\def\Fano{\mathcal{F}}
\def\Tw{T_{\mathrm{w}}}
\def\Bperp{B_{\bot}}
\def\kF{k_{F}}
\def\mfp{\ell}
\def\Rdiff{R}
\def\K{\mathrm{K}}
\def\mK{\mathrm{mK}}
\def\MHz{\mathrm{MHz}}
\def\mV{\mathrm{mV}}
\def\nA{\mathrm{nA}}
\def\uV{\mu\mathrm{V}}
\def\nm{\mathrm{nm}}
\def\um{\mu \mathrm{m}}
\def\kohm{\mathrm{k}\Omega}
\def\V{\mathrm{V}}
\def\pcmsq{\mathrm{cm}^{-2}}
\def\Tesla{\mathrm{T}}
\def\sigmin{\sigma_{\mathrm{min}}}
\def\Ohm{\Omega}

\begin{document}
\title{Shot Noise in Graphene }
\author{L.\ DiCarlo}
\affiliation{Department of Physics, Harvard University, Cambridge, MA 02138, USA}
\author{J.\ R.\ Williams}
\affiliation{School of Engineering and Applied Sciences, Harvard University, Cambridge, MA 02138, USA}
\author{Yiming\ Zhang}
\affiliation{Department of Physics, Harvard University, Cambridge, MA 02138, USA}
\author{D.\ T.\ McClure}
\affiliation{Department of Physics, Harvard University, Cambridge, MA 02138, USA}
\author{C.\ M.\ Marcus}
\affiliation{Department of Physics, Harvard University, Cambridge, MA 02138, USA}

\date{\today}

\begin{abstract}
We report measurements of current noise in single- and
multi-layer graphene devices. In four single-layer devices, including
a \emph{p-n} junction, the Fano factor remains constant to within $\pm10\%$ upon varying carrier type and
density, and averages between 0.35 and 0.38. The Fano factor in a multi-layer device is found to decrease from a maximal value of 0.33 at the charge-neutrality point to 0.25 at high carrier density. These results are compared to theories for shot noise in ballistic and disordered graphene.
\end{abstract}

\maketitle

Shot noise, the temporal fluctuation of electric current out of equilibrium, originates from the partial transmission of quantized charge~\cite{Blanter00}.
Mechanisms that can lead to shot noise in mesoscopic conductors include tunneling, quantum interference, and scattering from impurities and lattice defects. Shot noise yields information about transmission that is not available from the dc current alone.

In graphene~\cite{Geim07,CastroNeto07}, a zero-gap two-dimensional semi-metal in which carrier type and density can be controlled by gate voltages~\cite{Novoselov04},
density-dependent shot-noise signatures under various conditions have been investigated theoretically~\cite{Tworzydlo06,Cheianov06}. For wide samples of ballistic graphene (width-to-length ratio $W/L\gtrsim 4$) the Fano factor, $\Fano$, i.\,e., the current noise normalized to the noise of Poissonian transmission statistics, is predicted to be 1/3 at the charge-neutrality point and  $\sim 0.12$ in both electron (\emph{n}) and hole (\emph{p}) regimes ~\cite{Tworzydlo06}. The value $\Fano=1-1/\sqrt 2\approx 0.29$ is predicted for shot noise across a ballistic \emph{p-n} junction~\cite{Cheianov06}. For strong, smooth ``charge-puddle" disorder, theory predicts $\Fano \approx 0.30$ both at and away from the charge-neutrality point, for all $W/L \gtrsim 1$~\cite{SanJose07}.
Disorder may thus have a similar effect on noise in graphene as in diffusive metals, where $\Fano$ is universally $1/3$~\cite{Beenakker92,deJong92,Nazarov94,Steinbach96,Henny99,Schoelkopf97} regardless of shape and carrier density.
Recent theory investigates numerically the evolution from a density-dependent to a density-independent $\Fano$ with increasing disorder~\cite{Lewenkopf07}.
To our knowledge, experimental data for shot noise in graphene has not yet been reported.

This Letter presents an experimental study of shot
noise in graphene at low temperatures and zero magnetic field. Data for five devices, including a locally-gated \emph{p-n} junction, are presented. For
three globally-gated, single-layer samples, we find $\Fano \sim 0.35-0.37$ in both electron and hole doping regions, with essentially no dependence on electronic sheet density, $\den$, in the range $|\den| \lesssim 10^{12}~\pcmsq$.  Similar values are obtained for a
locally-gated single-layer \emph{p-n} junction
in both unipolar (\emph{n-n} or \emph{p-p}) and bipolar (\emph{p-n} or \emph{n-p}) regimes.
In a multi-layer sample, the observed $\Fano$ evolves from $0.33$ at the charge-neutrality point to $0.25$ at $\den \sim 6 \times 10^{12}~\pcmsq$.

\begin{figure}[t!]
\center \label{fig1}
\includegraphics[width=3.25in]{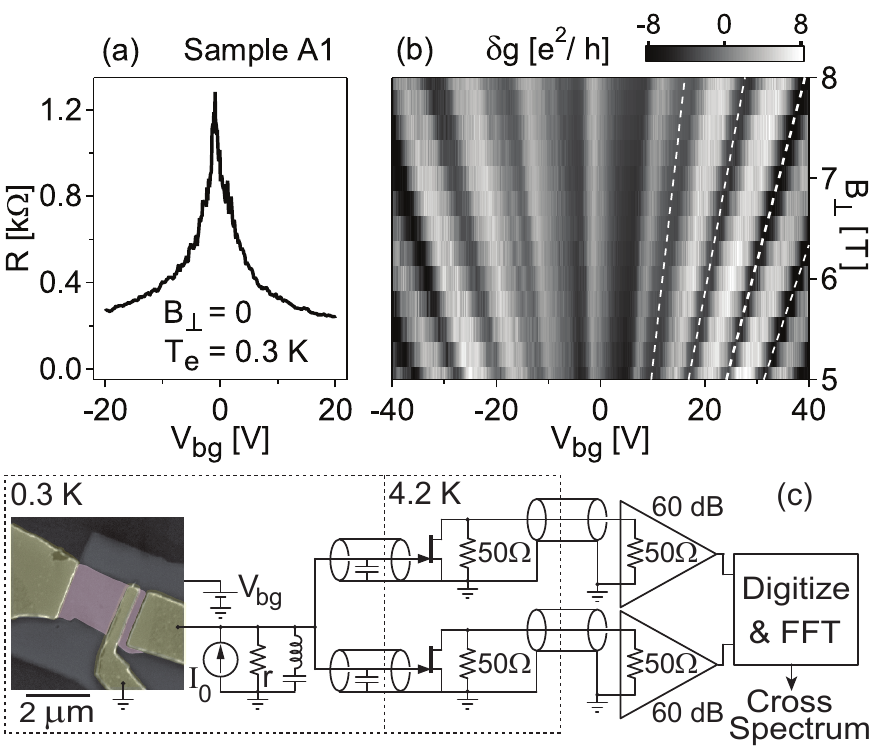}
\caption{\footnotesize{(a) Differential resistance $\Rdiff$ of sample
A1 as a function of back-gate voltage $\Vbg$ at electron temperature $\Te = 0.3~\K$, perpendicular field
$\Bperp=0$, and source-drain voltage $\Vsd=0$. (b) Differential two-terminal conductance $g(\Vsd=0)$ as a function of $\Bperp$ and
$\Vbg$ in the quantum Hall regime,after subtracting a quadratic fit at each $\Bperp$. Lines of constant filling factors 6, 10,
14, and 18 (dashed lines) indicate a single-layer sample.
(c) Equivalent circuit near $1.5~\MHz$ of the system measuring
current noise  using cross correlation of two channels~\cite{Techniques06}.
Current bias $I_o$ contains a
$7.5~\nA_{\mathrm{rms}}$, 20 Hz part for lock-in measurements and a
controllable dc part generating the dc component of $\Vsd$ via the shunt resistance $r=5~\kohm$.
False-color scanning electron micrograph of a three-lead pattern defining two devices similar to A1 and A2.
Purple indicates single-layer graphene and gold indicates metallic contacts.}}
\end{figure}

Devices were fabricated by mechanical exfoliation of highly-oriented
pyrolytic graphite~\cite{Novoselov04}. Exfoliated sheets were
deposited on a degenerately-doped Si substrate capped with $300~\nm$
of thermally grown $\SiOx$. Regions identified by optical microscopy
as potential single-layer graphene were contacted with
thermally evaporated Ti/Au leads (5/40~$\nm$) patterned by
electron-beam lithography. Additional steps in the fabrication of
the \mbox{\emph{p-n}} junction device are detailed in Ref.~\cite{Williams07}. Devices
were measured in two $\Hethree$ cryostats, one allowing
dc (lock-in) transport measurements in fields  $|\Bperp|\leq 8 ~\Tesla$ perpendicular to the graphene plane, and another
allowing simultaneous measurements of dc transport and noise~\cite{Techniques06} near
$1.5~\MHz$, but limited to $\Bperp\sim 0$.

Differential resistance  $\Rdiff=d\Vsd/d\Isd$ ($\Isd$ is the current,  and $\Vsd$ is the source-drain
voltage) of a wide, short sample [A1, $(W,L)=(2.0,0.35)~\um$]  is shown as a function of back-gate voltage $\Vbg$ at $\Vsd=0$ and $\Bperp=0$ in Fig.~1(a).
While the width of the peak is consistent with A1 being
single-layer graphene~\cite{Novoselov05,Zhang05}, more direct evidence is
obtained from the QH signature shown in Fig.~1(b). The grayscale
image shows differential conductance $g=1/\Rdiff$ as a function of $\Vbg$
and $\Bperp$, following subtraction of the best-fit quadratic polynomial to $g(\Vbg)$ at
each $\Bperp$ setting to maximize contrast. Dashed lines correspond
to filling factors $\den h/e\Bperp=6$,  10, 14, and 18, with
$\den=\alpha(\Vbg+1.1~\V)$ and lever arm $\alpha=6.7\times
10^{10}\ \pcmsq/\V$. Their alignment with local minima in
$\delta g(\Vbg)$ identifies A1 as single-layer
graphene~\cite{McCann06, Martin07}. The Drude mean free path
$\mfp=\Rq \cdot \sigma/\kF$~\cite{Rycerz06}, where $\kF=\sqrt{\pi |\den|}$,
is found to be $\sim 40~\nm$ away from the charge-neutrality point using the
$\Bperp=0$ conductivity $\sigma=(\Rdiff W/L)^{-1}$ [Fig.~2(a) inset].

\begin{figure}[t!]
\center \label{fig2}
\includegraphics[width=3.25in]{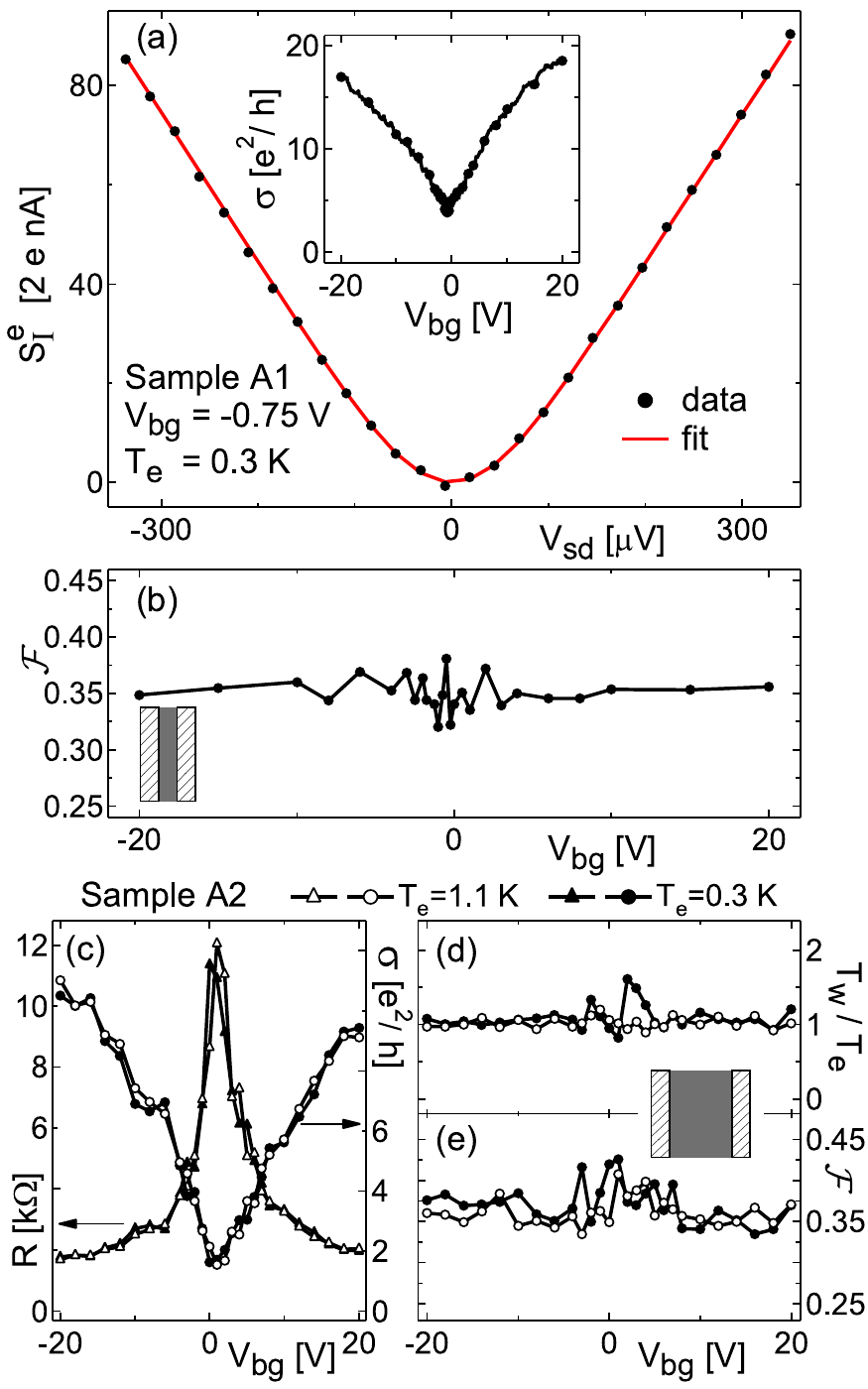}
\caption{\footnotesize{(a) Inset: Conductivity
$\sigma=(\Rdiff W/L)^{-1}$ calculated using $\Rdiff(\Vbg)$ data in Fig.~1(a) and
$W/L=5.7$.  Solid black circles correspond to
$\sigma(\Vsd=0)$ at the $\Vbg$ settings of noise measurements shown in (b).
Main: Excess noise $\Siex$ as function of $\Vsd$ near the charge-neutrality
point, $\Vbg=-0.75~\V$. The solid red curve is the
single-parameter best fit to Eq.~(1), giving Fano factor
$\Fano=0.349$ (using $\Te=303~\mK$ as calibrated by JNT). (b) Best-fit
$\Fano$ at 25 $\Vbg$ settings across the charge-neutrality point for
electron and hole densities reaching $|\den| \sim1.4\times10^{12}~\pcmsq$.
(c) $\Rdiff$ (left axis) and $\sigma$ (right axis) of sample A2 as a function of $\Vbg$
$(W/L=1.4)$, with $\Vsd=0$, at $0.3~\K$ (solid markers) and at $1.1~\K$ (open
markers). (d), (e) Crossover width $\Tw$ (normalized to
JNT-calibrated $\Te$) and $\Fano$, obtained from best-fits
using Eq.~(1) to $\Siex(\Vsd)$ data over $|\Vsd|\leq 350(650)~\uV$
for $\Te=0.3 (1.1)~\K$.}}
\end{figure}

Current noise spectral density $S_I$ is measured
using a cross-correlation technique described in Ref.~\cite{Techniques06} [see Fig.~1(c)].  Following
calibration of amplifier gains and electron temperature $\Te$ using
Johnson noise thermometry (JNT) for each
cooldown, the excess noise $\Siex\equiv S_I - 4 \Kb \Te g (\Vsd)$ is
extracted. $\Siex(\Vsd)$ for sample A1 is shown in Fig.~2(a). Linearity of $\Siex$ at high bias indicates negligible extrinsic ($1/f$ or telegraph) resistance fluctuations within the measurement bandwidth.
For these data, a single-parameter fit to the scattering-theory form (for energy-independent transmission) ~\cite{Lesovik89,Buttiker92},
\begin{equation}
\Siex=2e \Isd \mathcal{F}\left [\coth\left(\frac{e\Vsd}{2 \Kb
\Te}\right)- \frac{2\Kb\Te}{e\Vsd}\right],
\end{equation}
gives a  best-fit Fano factor $\Fano=0.349$.
Simultaneously measured conductance $g\approx
22.2~\Gq$ was independent of bias within $\pm0.5\%$ (not shown) in the $|\Vsd| \leq 350~\uV$ range used for the fit. Note that the observed
quadratic-to-linear crossover agrees well with that in the curve fit, indicating weak inelastic scattering in A1~\cite{Steinbach96,Henny99}, and negligible series resistance (e.\,g., from contacts), which would broaden the crossover by reducing the effective $\Vsd$ across the sample.

Figure~2(b) shows similarly measured values for
$\Fano$ as a function of $\Vbg$. $\Fano$ is observed to remain nearly constant for $|\den| \lesssim
10^{12}~\pcmsq$. Over this density range, the average $\Fano$ is $0.35$ with standard deviation $0.01$.
The estimated error in the best-fit $\Fano$ at each $\Vbg$ setting is $\pm 0.002$, comparable to
the marker size and smaller than the variation in $\Fano$ near $\Vbg=0$, which we believe results from mesoscopic
fluctuations of $\Fano$. Nearly identical noise results (not shown) were found for a similar sample (B), with
dimensions $(2.0,0.3)~\um$ and a QH signature consistent with a single layer.

Transport and noise data for a more square single-layer sample [A2, patterned on the same graphene sheet as A1, with dimensions $(1.8,1.3)~\um$] at $\Te=0.3~\K$ (solid circles) and $\Te=1.1~\K$ (open circles) are shown in Figs.~2(c-e). At both temperatures,
the conductivity shows $\sigmin\approx 1.5~\Gq$ and gives $\mfp\sim 25~\nm$ away from the charge-neutrality point. That these two values differ from those in sample A1 is particularly notable as
samples A1 and A2 were patterned on the same piece of graphene. Results of fitting Eq.~(1) to $\Siex(\Vsd)$ for sample A2
are shown in Figs.~2(d) and 2(e). To allow for possible broadening
of the quadratic-to-linear crossover by series resistance and/or
inelastic scattering, we treat electron temperature as a second fit parameter
(along with $\Fano$) and compare the best-fit value, $\Tw$, with the
$\Te$ obtained from Johnson noise. Figure~2(d) shows $\Tw$ tracking the calibrated
$\Te$ at both temperatures. Small deviation of  $\Tw/\Te$ from unity near the
charge-neutrality point at $\Te=0.3~\K$ can be attributed to
conductance variations up to $\pm20\%$ in the fit range $|\Vsd|\leq
350~\uV$ at these values of $\Vbg$. As in sample A1, $\Fano$ is found to be independent of carrier type and density over  $|\den|\lesssim 10^{12}~\pcmsq$, averaging 0.37(0.36) with standard deviation 0.02(0.02) at $T_e=0.3(1.1)$~K. Evidently, despite its
different aspect ratio, A2 exhibits a noise signature similar to that of A1.

The lack of $R$-dependence in $\Fano$  suggests that bias-dependent electron heating in the metallic reservoirs~\cite{Henny99} is negligible for our samples.
This heating, originating from imperfect dissipation of the generated power $\Vsd^2/R$, can affect shot noise measurements since these
require $|\Vsd|$ several times the thermal voltage (here, $e |\Vsd|/\Kb \Te \lesssim 10$). In the presence of heating,
fitting the excess noise $\Si(\Vsd,\Te+\Delta\Te(\Vsd))- 4 \Kb \Te g$ to Eq.~(1) overestimates $\Fano$.
The nearly equal values of $\Fano$ observed in A1 and A2 despite the factor $\sim 10$ difference in $R$ at comparable $\den$
suggest that heating in the reservoirs is negligible~\cite{HeatingNote}.

Transport and noise measurements for a single-layer
graphene \emph{p-n} junction~\cite{Williams07}, sample C, are shown in Fig.~3. The color image
in Fig.~3(a) shows differential resistance $\Rdiff$ as a function of
$\Vbg$ and local top-gate voltage $\Vtg$. The two gates allow independent control
of charge densities in adjacent regions of the device [see Fig.~3(c) inset].
In the bipolar regime, the best-fit $\Fano$ shows little density dependence and averages 0.38,
equal to the average value deep in the unipolar regime, and similar to results for the back-gate-only single-layer samples (A1, A2 and B). Close to charge neutrality in either region (though particularly in the region under the top gate), $\Siex(\Vsd)$ deviates from the form of Eq.~(1) (data not shown). This is presumably due to resistance fluctuation near charge neutrality, probably due mostly to mobile traps in the $\AlOx$ insulator beneath the top gate.

\begin{figure}[t!]
\center \label{fig3}
\includegraphics[width=3.25in]{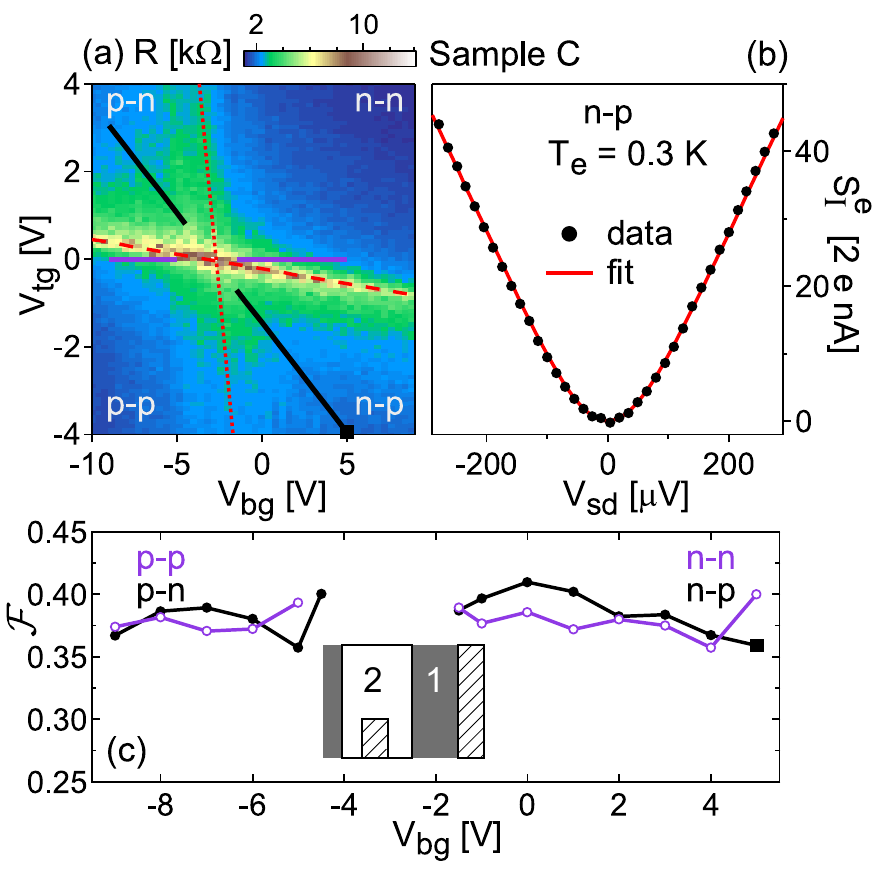}
\caption{\footnotesize{(a) Differential resistance $\Rdiff$ of sample
C, a single-layer \emph{p-n} junction, as a function of back-gate voltage
$\Vbg$ and top-gate voltage $\Vtg$. The skewed-cross pattern defines
quadrants of $n$ and $p$ carriers in regions 1 and 2. Red lines
indicate charge-neutrality lines in region 1 (dotted) and region 2 (dashed). (b)
$\Siex(\Vsd)$ measured in \emph{n-p} regime with
$(\Vbg,\Vtg)=(5,-4)~\V$ (solid dots) and best fit to Eq.~(1)
(red curve), with $\Fano=0.36$. (c) Main: Best-fit $\Fano$ along
the cuts shown in (a), at which  $n_{s1} \sim n_{s2}$ (purple)
and $n_{s1} \sim- 4~n_{s2}$ (black).
Inset: Schematic of the device. The top gate covers region 2 and one
of the contacts.}}
\end{figure}

Measurements at $0.3~\K$ and at $1.1~\K$ for sample D, of dimensions $(1.8,1.0)~\um$,
are shown in Fig.~4. A $\sim~3~\nm$ step height between $\SiOx$ and carbon surfaces
measured by atomic force microscopy prior to electron-beam lithography~\cite{Graf07} suggests this
device is likely multi-layer. Further indications include  the broad $\Rdiff(\Vbg)$ peak~\cite{Zhang05a} and the large minimum conductivity, $\sigmin\sim 8~\Gq$ at $\Bperp=0$ [Fig.~4(a)], as well as the absence of QH signature for $|\Bperp|\leq 8~\Tesla$ at $250~\mK$ (not shown). Two-parameter fits of $\Siex(\Vsd)$ data to Eq.~(1) show three notable differences from results in the single-layer samples [Figs.~4(b) and 4(c)]: First, $\Fano$ shows a measurable dependence on back-gate voltage, decreasing from 0.33 at the charge-neutrality point to 0.25 at $\den\sim 6\times10^{12}~\pcmsq$ for  $\Te=0.3~\K$;
Second, $\Fano$ decreases with increasing temperature; Finally,  $\Tw/\Te$ is 1.3-1.6 instead of very close to 1. We interpret the last two differences, as well as the sublinear dependence of $\Siex$ on $\Vsd$ (see Fig.~4 inset) as indicating sizable inelastic scattering~\cite{Beenakker92,deJong92} in sample D.
(An alternative explanation in terms of series resistance would require it to be density, bias, and temperature dependent, which is inconsistent with the independence of $g$ on $\Vsd$ and $\Te$).

\begin{figure}[t!]
\center \label{fig4}
\includegraphics[width=3.25in]{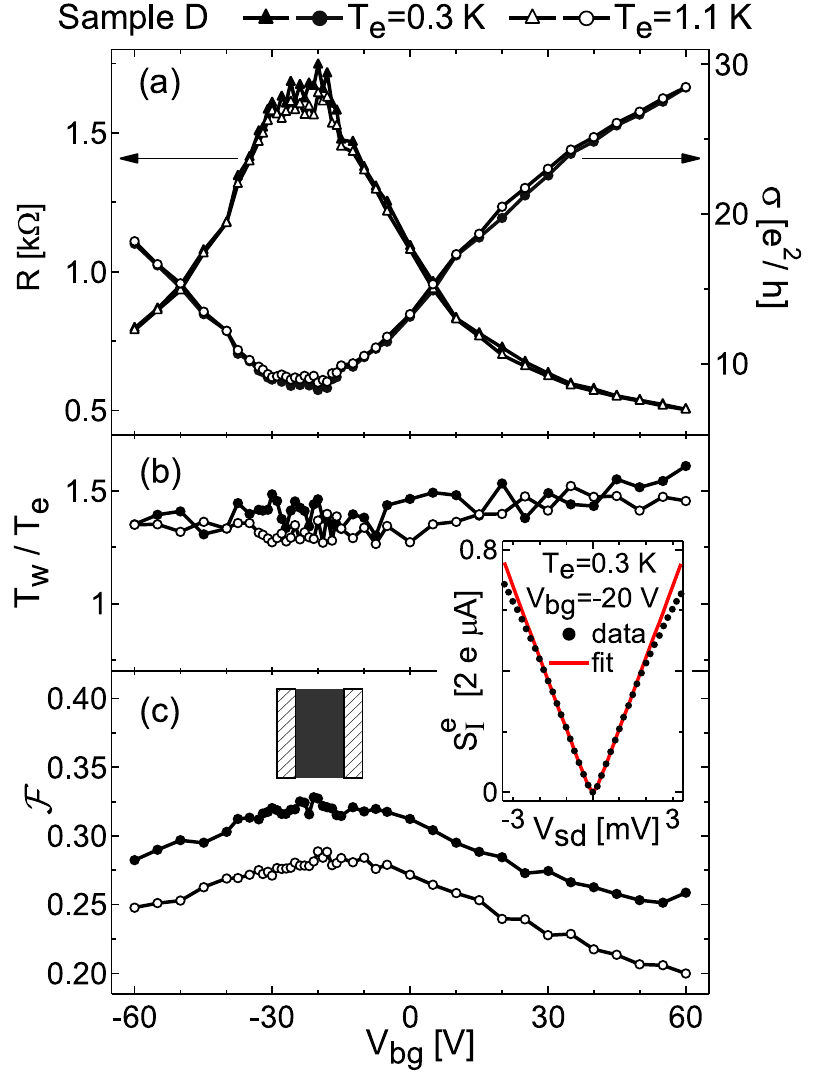}
\caption{\footnotesize{(color) (a) Differential resistance $\Rdiff$ (left axis) and
conductivity $\sigma$ (right axis) of sample D as a function of
$\Vbg$, with $\Vsd=0$, at $0.3~\K$ (solid markers) and at $1.1~\K$ (open
markers). (b),(c) Best-fit $\Tw$ (normalized to
JNT-calibrated $\Te$) and $\Fano$ to $\Siex(\Vsd)$ data over
$|\Vsd|\leq 0.5(1)~\mV$ for $\Te=0.3 (1.1)~\K$. Inset: Sublinear dependence of $\Siex$ on $\Vsd$ is evident in data taken over a larger bias range.
Solid red curve is the two-parameter best fit of Eq.~(1) over $|\Vsd|\leq 0.5~\mV$.}}
\end{figure}

Summarizing the experimental results, we find that in four single-layer samples, $\Fano$ is insensitive to
carrier type and density, temperature, aspect ratio, and the presence of a \emph{p-n} junction.
In one multi-layer sample, $\Fano$ does depend on density and temperature,
and $\Siex(\Vsd)$ shows a broadened quadratic-to-linear crossover and is sublinear in $\Vsd$ at high bias. We may now compare these results to expectations based on theoretical and numerical results for ballistic and disordered graphene.

Theory for ballistic single-layer graphene with $W/L~\gtrsim~4$ gives a universal
$\Fano=1/3$ at the charge-neutrality point, where transmission is evanescent, and $\Fano\sim 0.12$
for $|\den| \gtrsim \pi/L^2$, where propagating modes dominate transmission~\cite{Tworzydlo06}.
While the measured $\Fano$ at the charge-neutrality point in samples A1 and B ($W/L=5.7$ and $6.7$, respectively) is consistent with this prediction, the absence of density dependence is not: $\pi/L^2 \sim 3 \times 10^9~\pcmsq$ is well within the range of carrier densities covered in the measurements.
Theory~\cite{Schomerus07} for ballistic graphene contacted with finite-density leads finds slight increments of $\Fano$ from 1/3 at the charge-neutrality point, in agreement with this experiment. However, $\Fano$ in this contact model remains density dependent.
Theory for ballistic graphene \emph{p-n} junctions \cite{Cheianov06} predicts $\Fano\approx 0.29$, lower than the value $\sim 0.38$ observed
in sample C in both \emph{p-n} and \emph{n-p} regimes. We speculate that these discrepancies likely arise from the presence of disorder. Numerical results for strong, smooth disorder \cite{SanJose07} predict a constant $\Fano$  at and away from the charge-neutrality point for $W/L \gtrsim 1$, consistent with experiment.
However, the predicted value $\Fano\approx 0.30$ is $\sim 20\%$ lower than observed in all single-layer devices.
Recent numerical simulations \cite{Lewenkopf07} of small samples ($L=W\sim 10~\nm$) investigate the vanishing of carrier dependence in $\Fano$ with increasing disorder strength. In the regime where disorder makes $\Fano$ density-independent, the value $\Fano \sim0.35-0.40$ is found to depend weakly on disorder strength and sample size.

Since theory for an arbitrary number of layers is not available for comparison to noise results in the multi-layer
sample D, we compare only to existing theory for ballistic bi-layer graphene~\cite{Snyman07}.
It predicts $\Fano=1/3$ over a much narrower density range than for the single layer,
and abrupt features in $\Fano$ at finite density due to transmission resonances.
A noise theory beyond the bi-layer ballistic regime may thus be necessary to explain the observed smooth decrease of $\Fano$ with increasing density in sample D.

We thank C.~H. Lewenkopf, L.~S. Levitov, and D.~A. Abanin for useful
discussions. Research supported in part by the IBM Ph.D. Fellowship program (L.D.C.), INDEX, an NRI Center, and
Harvard NSEC.


\begin{thebibliography}{10}
\bibitem{HeatingNote}
Effects of reservoir heating on noise have been observed in diffusive wires ($R\sim 10~\Ohm$)
with metallic contacts~\cite{Henny99}. The generated power in our graphene samples is weaker by 1-3 orders of magnitude,
which could explain why heating of the metallic contacts appears negligible in this experiment.
\end{thebibliography}
\end{document}